\documentstyle[12pt]{article}
\newcommand{\EQ}{\begin{equation}}
\newcommand{\EN}{\end{equation}}
\newcommand{\bea}{\begin{eqnarray}}
\newcommand{\eea}{\end{eqnarray}}
\newcommand{\hs}{\hspace{0.1cm}}

\newcommand{\st}{\stackrel}
\newcommand{\th}{\theta}
\newcommand{\var}{\varepsilon}
\newcommand{\al}{\alpha}

\newcommand{\goto}{\rightarrow}
\newcommand{\lab}{\label}

\begin{document}
\setcounter{page}{0}
\topmargin 0pt
\oddsidemargin 5mm
\renewcommand{\thefootnote}{\arabic{footnote}}
\newpage
\setcounter{page}{0}
\begin{titlepage}
\begin{flushright}
OUTP-96-19S\\
SWAT/95-96/110
\end{flushright}
\vspace{0.5cm}
\begin{center}
{\large {\bf Correlation Functions in the Two-Dimensional \\
Ising Model in a Magnetic Field at $T=T_{c}$}}\\
\vspace{1.8cm}
{\large G. Delfino} \\
\vspace{0.5cm}
{\em Theoretical Physics, University of Oxford}\\
{\em 1 Keble Road, Oxford OX1 3NP, United Kingdom}\\
{\em email: g.delfino1@physics.oxford.ac.uk}\\
\vspace{0.5cm}
{\large and}\\
\vspace{0.5cm}
{\large P. Simonetti}\\
\vspace{0.5cm}
{\em Department of Physics, University of Wales Swansea,\\
Singleton Park, Swansea SA2 8PP, United Kingdom}\\
{\em email: p.simonetti@swansea.ac.uk}\\
\end{center}
\vspace{1.2cm}

\renewcommand{\thefootnote}{\arabic{footnote}}
\setcounter{footnote}{0}

\begin{abstract}
\noindent
The one and two-particle form factors of the energy operator in the
two-dimensional Ising model in a magnetic field at $T=T_c$ are exactly
computed within the form factor bootstrap approach. Together with the
matrix elements of the magnetisation operator already
computed in ref.\,\cite{immf}, they are used to write down the large
distance expansion for the correlators of the two relevant fields of the
model.  
\end{abstract}

\vspace{.3cm}

\end{titlepage}

\newpage
\noindent
The last years have seen important progresses in the non-perturbative
study of two-dimensional quantum field theories and related
statistical mechanical models. If conformal symmetry provided us with
an exact description of critical points and universality classes
\cite{BPZ,ISZ}, the study of off-critical models turned out to be
better approached within the framework of relativistic scattering
theory. In fact, if the off-critical model under consideration is
integrable (i.e. admits infinite conservation laws), it can be usually
solved exploiting very general bootstrap techniques
\cite{ZZ,Zam,GM,Smirnov}. This circumstance appears to be particularly
important in light of the fact that a large number of physically
interesting two-dimensional systems can actually be described in terms
of integrable models. A remarkable example is provided by
the scaling limit of the two-dimensional Ising model in a magnetic
field at $T=T_c$ (IMMF in the sequel) \cite{Zam}. It can be 
formally described by the action
\EQ
{\cal A} = {\cal A}_{CFT} + h \int d^2x \, \sigma(x) \,\,,
\label{action}
\EN
where ${\cal A}_{CFT}$ denotes the action of the conformal minimal
model ${\cal M}_{3,4}$ and $\sigma(x)$ the magnetisation operator of
scaling dimension $2\Delta^\sigma=1/8$. The coupling constant $h$
(magnetic field) has physical dimension $h\sim m^{15/8}$, $m$ being a
mass scale. Apart form the magnetisation operator, the only other
relevant scaling field in the Ising model is the energy density
$\var(x)$ with scaling dimension $2\Delta^\var=1$.

Zamolodchikov showed that the theory described by the
action (\ref{action}) possesses an infinite number of integrals of
motion which can be used in order to determine the exact particle
spectrum and $S$--matrix of the theory \cite{Zam}. He found that the
spectrum consists of eight massive particles $A_a$ ($a=1,2,\ldots,8$)
whose masses stay in the following ratios with the mass $m_1$ of the
lightest particle
\bea
m_2 &=& 2 m_1 \cos\frac{\pi}{5} = (1.6180339887..) \,m_1\,,\nonumber\\
m_3 &=& 2 m_1 \cos\frac{\pi}{30} = (1.9890437907..) \,m_1\,,\nonumber\\
m_4 &=& 2 m_2 \cos\frac{7\pi}{30} = (2.4048671724..) \,m_1\,,\nonumber \\
m_5 &=& 2 m_2 \cos\frac{2\pi}{15} = (2.9562952015..) \,m_1\,,\\
m_6 &=& 2 m_2 \cos\frac{\pi}{30} = (3.2183404585..) \,m_1\,,\nonumber\\
m_7 &=& 4 m_2 \cos\frac{\pi}{5}\cos\frac{7\pi}{30} = (3.8911568233..) \,m_1\,,
\nonumber\\
m_8 &=& 4 m_2 \cos\frac{\pi}{5}\cos\frac{2\pi}{15} = (4.7833861168..) \,m_1\,.
\nonumber
\eea
The interaction between these particles is described by a factorised, 
reflectionless $S$--matrix characterised by the two--particle amplitudes
\EQ
S_{ab}(\th) = \prod_{\alpha\in{\cal A}_{ab}}
\left[
\frac{\tanh\frac{1}{2}\left(\th+i\pi
\al\right)}
     {\tanh\frac{1}{2}\left(\th-i\pi \al\right)}
\right]^{\mu_\alpha}\,\,\,.
\label{ampl}
\EN
The set of numbers ${\cal A}_{ab}$ and the multiplicity factors
$\mu_\alpha$ are given in Table\,1. We use the standard rapidity
parameterisation of the on--shell momenta
$p_a^\mu=(m_a\cosh\th_a,m_a\sinh\th_a)$, so that $\th\equiv\th_a-\th_b$
in (\ref{ampl}).

In ref.\,\cite{immf} the knowledge of the $S$--matrix (\ref{ampl}) was
exploited in order to approach the computation of the correlation
functions of the model (\ref{action}) within the form factor bootstrap
method. The basic idea of this approach is to express the (euclidean)
correlation functions (e.g. the two--point ones) as a spectral sum
over a complete set of intermediate multiparticle states
%\footnote{See ref.\,\cite{qism} for an alternative approach
%to the computation of correlation functions in integrable models.}
\begin{eqnarray}
G_{\Phi_1\Phi_2}(x)&\equiv&\langle\Phi_1(x) \Phi_2(0)\rangle\nonumber\\
& = & \sum_{n=0}^{\infty}\int_{\th_1 >\th_2 \ldots>\th_n} 
\frac{d\th_1}{2\pi} \cdots \frac{d\th_n}{2\pi}\,\,
e^{-|x| \sum_{k=1}^n m_k \cosh\th_k}
\label{spec} \\
& \times & \langle 0|\Phi_1(0)|A_{a_1}(\th_1) \ldots A_{a_n}(\th_n)\rangle
\langle A_{a_1}(\th_1) \ldots A_{a_n}(\th_n)|\Phi_2(0)|0\rangle\,\,,\nonumber
\end{eqnarray}
and to exploit the fact that the form factors (FF)
\EQ
F^{\Phi}_{a_1\ldots a_n}(\th_1,\ldots,\th_n) = \langle 0|
\Phi(0)|A_{a_1}(\th_1)\ldots A_{a_n}(\th_n)\rangle\,\,\,.
\label{form}
\EN
are exactly computable in integrable models once the $S$--matrix is
known \cite{KW,Smirnov}.

The spectral series (\ref{spec}) is manifestly a large distance
expansion. Nevertheless, it has been observed in several models
[9--13] that it is characterised by a fast rate of convergence also
at intermediate and short distance scales (see in particular
ref.\,\cite{CMpol} for a theoretical justification of this property), so
that a truncation of the series including few lowest states turns out
to be sufficient
for most practical purposes. In ref.\,\cite{immf} the one and
two--particle FF of the magnetisation operator in the IMMF
were exactly computed and shown to be sufficient to reproduce with
remarkable accuracy the numerical data for $G_{\sigma\sigma}(x)$
avalaible from Monte Carlo simulations \cite{R-D}. It is the purpose
of this letter to carry out a similar program for the FF of the energy
operator $\varepsilon(x)$ 
and to write down the large distance expansion for the correlators
$G_{\var\var}(x)$ and $G_{\sigma\var}(x)$.

Let's briefly recall the basic strategy for the computation of FF
in the IMMF referring the reader to ref.\,\cite{immf} for
details. Form factors can be generally computed in an integrable model
using a recursive procedure based on a set of residue equations
relating matrix elements with different particle content. For
instance, if the scattering amplitude $S_{ab}(\th)$ has a simple pole
with positive residue $\left(\Gamma_{ab}^c\right)^2$ at
$\th=iu_{ab}^c$ corresponding to the particle $A_c$ with mass
$m_c^2=m_a^2+m_b^2+2m_am_b\cos u_{ab}^c$ appearing as a bound state in
the direct $ab$ channel, then we can write
\EQ
F^{\Phi}_{ab}(\th\simeq
iu_{ab}^c)\simeq\frac{i \Gamma_{ab}^c}{\th-iu_{ab}^c}F^{\Phi}_c\,\,,
\lab{pole}
\EN
as well as similar relations among higher matrix elements containing
spectator particles. Other recursive equations are associated to the
higher poles in the $S$--matrix and to the ``kinematical'' poles in
the matrix elements. Together with the equations ruling the monodromy
properties of FF, these residue equations provide a system of {\em
linear} relations whose general solution amounts to a complete
classification of the operator content of the theory \cite{JLG}.
 
In theories with diagonal scattering, the basic
information about the structure of FF is already encoded in the
two--particle matrix elements. In the IMMF they can be generally
parameterised as \cite{immf}
\footnote{In this letter we consider only scalar operators.}
\EQ
F^{\Phi}_{ab}(\th)=\frac{Q^{\Phi}_{ab}(\th)}{D_{ab}(\th)}
F^{min}_{ab}(\th)\,\,,
\lab{param}
\EN
where 
\EQ
F^{min}_{ab}(\th)=\left(-i\sinh\frac{\th}{2}\right)^{\delta_{ab}}
\prod_{\alpha\in{\cal A}_{ab}}\left(G_{\alpha}(\th)
\right)^{\mu_\alpha}\,\,,
\lab{fmin}
\EN
\EQ
G_{\al}(\th)=\exp\left\{2\int_0^\infty\frac{dt}{t}\frac{\cosh\left(
\al - \frac{1}{2}\right)t}{\cosh\frac{t}{2}\sinh
t}\sin^2\frac{(i\pi-\th)t}{2\pi}\right\}\,\,,
\lab{block}
\EN
and
\EQ
D_{ab}(\th)=\prod_{\alpha\in {\cal
A}_{ab}} \left({\cal P}_\alpha(\th)\right)^{i_\alpha}
\left({\cal P}_{1-\alpha}(\th)\right)^{j_\alpha} \,\,\,,
\lab{dab}
\EN
\EQ
\begin{array}{lll}
i_{\alpha} = n+1\,\,\, , & j_{\alpha} = n \,\,\, , &
{\rm if} \hspace{.5cm} \mu_\alpha=2n+1\,\,\,; \\
i_{\alpha} = n \,\,\, , & j_{\alpha} = n \,\,\, , &
{\rm if} \hspace{.5cm} \mu_\alpha=2n\,\,\, ,
\end{array}
\EN
\EQ
{\cal P}_{\al}(\th)\equiv
\frac{\cos\pi\al-\cosh\th}{2\cos^2\frac{\pi\al}{2}}\,\,\,.
\lab{polo}
\EN
The terms $F_{ab}^{min}(\th)$ and $D_{ab}(\th)$ in eq.\,(\ref{param})
take into account the monodromy properties and the singularity
structure of the matrix elements, respectively, and are both uniquely
determined by the knowledge of the $S$--matrix. The whole information
about the operator $\Phi(x)$ is then contained in the polynomial
\EQ
Q_{ab}^\Phi(\th)=\sum_{k=0}^{N_{ab}} c^k_{ab}\,\cosh^k\th\,\,,
\EN
whose (operator dependent) coefficients $c_{ab}^k$ are the only
remaining unknowns. 

The identification of specific operators out of the general solution
of the recursive equations is a nontrivial task, especially in
theories lacking any internal symmetry, as is the case for the IMMF. A
progress with respect to this problem was made in ref.\,\cite{immf}
where it was shown that in a unitary theory the FF of an
operator $\Phi(x)$ satisfy the asymptotic bound
\EQ
\lim_{|\th_i|\goto\infty}
F^{\Phi}_{a_1\ldots a_n}(\th_1,\ldots,\th_n)\leq const.\,
e^{\Delta_{\Phi}|\th_i|}\,\,,
\lab{bound}
\EN
$2\Delta^\Phi$ being the scaling dimension. The consequences of this
result are easily illustrated using the parameterisation (\ref{param}).
Consider the simplest two-particle FF in the IMMF, $F_{11}^\Phi(\th)$. 
Since $G_{\al} (\th) \,\sim\, \exp(|\th|/2)$ as $|\th|\goto\infty$, the
constraint (\ref{bound}) implies that, for any relevant scalar field
$\varphi(x)$ ($\Delta_\varphi<1$), the total degree $N_{11}$ of the polynomial
$Q_{11}^\varphi(\th)$ must be less than 2. Since it can be checked
that no solution of the residue equations exists if $N_{11}=0$, one
concludes
\EQ
Q_{11}^\varphi(\th)=c_{11}^1\cosh\th+c_{11}^0\,,\hspace{1cm}c_{11}^1\neq 0\,\,.
\EN
It turns out that, once the FF $F_{11}^\varphi(\th)$ corresponding to
a specific relevant operator $\varphi(x)$ has been fixed assigning the
two coefficients $c_{11}^1$ and $c_{11}^0$, all the FF of $\varphi(x)$
can be uniquely determined using the residue equations. This amounts
to say that the solutions of the FF bootstrap for the relevant scalar
operators of the IMMF form a two--dimensional linear space, which is
what expected from the fact that such operators can only correspond to
linear combinations of $\sigma(x)$ and $\var(x)$. 
Solutions corresponding to operators not just differing for an
inessential normalisation constant can be labelled by the ratio
\EQ
z_\varphi\equiv \frac{c_{11}^0}{c_{11}^1}\,\,. 
\EN
It is of particular
physical interest to determine the values of this ratio which select
the two scaling fields $\sigma(x)$ and $\var(x)$. The problem for
$\sigma(x)$ was solved in ref.\,\cite{immf} using the relation between
the field perturbing the conformal point and the trace of the
energy--momentum tensor
\EQ
\Theta(x)=2\pi h\,(2-2\Delta^\sigma)\sigma(x)\,\,,
\EN
and exploiting the constraints imposed on the matrix elements of
$\Theta(x)$ by energy--momentum conservation. On the other hand, no
similar method can be used for $\var(x)$ and an alternative way must
be found in order to characterise the FF of this operator.

The fields $\sigma(x)$ and $\var(x)$ are
uniquely identified by the short distance behaviour of their
correlation functions predicted by the conformal operator product 
expansion \cite{BPZ}
\bea
G_{\sigma\sigma}(x) &\sim&
|x|^{-4\Delta_\sigma}=|x|^{-1/4}\,,\hspace{1cm}|x|\goto 0\nonumber\\
G_{\sigma\var}(x) &\sim&
|x|^{-2\Delta_\var}=|x|^{-1}\,,\hspace{1cm}|x|\goto 0 \lab{uv}\\
G_{\var\var}(x) &\sim&
|x|^{-4\Delta_\var}=|x|^{-2}\,,\hspace{1cm}|x|\goto 0\nonumber\,\,\,.
\eea
Since in the FF approach the determination of the exact ultraviolet
behaviour of the correlators requires in principle the resummation of
the spectral series (\ref{spec}), it seems quite difficult to make a 
direct use of eqs.\,(\ref{uv}) for the determination of $z_\sigma$ and
$z_\var$. Nevertheless, eqs.\,(\ref{uv}) suggest that, if a property
characterising the FF of the scaling fields $\sigma(x)$ and $\var(x)$
exists, it is probably related to the high energy asymptotics of the
matrix elements. Moreover, if such property should be able to select
the scaling fields among their linear combinations, it must be {\em non-linear}
in the operator. Interestingly enough, a property with these features
is known in the FF literature. It was noticed in the context of the
sine--Gordon model that the FF of the operators exponential
of the elementary field satisfy the {\em cluster property}
\cite{Smirnov,SmirnovSG,MS-KM}
\EQ
\lim_{\alpha\goto\infty}\tilde F_{a_1\ldots a_ka_{k+1}\ldots
a_n}^\Phi(\th_1+\alpha,\ldots,\th_k+\alpha,\th_{k+1},\ldots,\th_n)
=\tilde F^\Phi_{a_1\ldots a_k}(\th_1,\ldots,\th_k)
 \tilde F^\Phi_{a_{k+1}\ldots a_n}(\th_{k+1},\dots,\th_n)\,\,,
\lab{cluster}
\EN
where
\EQ
\tilde F^\Phi_{a_1\ldots a_n}(\th_1,\ldots,\th_n)\equiv
\frac{1}{\langle\Phi\rangle}
F^\Phi_{a_1\ldots a_n}(\th_1,\ldots,\th_n)\,\,.
\EN
It is known that massive deformations of minimal models can be
obtained from sine--Gordon through a suitable restriction of the
Hilbert space and that some exponential operators are mapped into the
scaling fields of the restricted models. In refs.\,\cite{SmirnovSG,Koubek}, 
it was shown for some specific cases that the factorisation property
(\ref{cluster}) survives the reduction procedure and is satisfied by
the FF of the scaling fields in the reduced models. Here we simply 
assume that the cluster
property (\ref{cluster}) characterises the FF of the operators
$\sigma(x)$ and $\var(x)$ in the IMMF and provide what we think is
strong evidence that this is indeed the case.

Before proceeding further, notice that
the asymptotic relation (\ref{cluster}) involves the vev $\langle
\Phi\rangle$. In the general case, the computation of this
quantity is a nontrivial problem (the thermodynamic Bethe ansatz
only provides the vev of the field which perturbs the conformal
point \cite{TBA}). It is then remarkable that, if the FF 
of $\Phi(x)$ satisfy the
cluster property, the vev can be obtained, for instance, as
\EQ
\langle\Phi\rangle=\frac{F_a^\Phi F_b^\Phi}{\lim_{\th\goto\infty}
F_{ab}^\Phi(\th)} \,\,.
\lab{vev}
\EN
Going back to the determination of $z_\sigma$ and $z_\var$,
consider the two--particle FF $F_{12}^\varphi(\th)$. Following the
same arguments used above for $F_{11}^\varphi(\th)$ one concludes that
$Q_{12}^\varphi(\th)$ is a polynomial of degree 2 in $\cosh\th$. 
Since the amplitudes $S_{11}(\th)$ and
$S_{12}(\th)$ have common poles corresponding to the particles $A_1$,
$A_2$ and $A_3$ (see Table\,1), eq.\,(\ref{pole}) provides the linear system 
\EQ
\frac{1}{\Gamma_{11}^c} {\rm Res}_{\th=iu_{11}^c}F_{11}^{\varphi}(\th)=
\frac{1}{\Gamma_{12}^c} {\rm Res}_{\th=iu_{12}^c}F_{12}^{\varphi}(\th)\,\,,
\hspace{1cm}c=1,2,3\,\,.
\EN
Once an overall normalisation of the operator $\varphi(x)$ has been
fixed, these equations
uniquely determine the coefficients $c_{12}^2$, $c_{12}^1$ and
$c_{12}^0$ in terms of $z_\varphi$. Finally, in order to search for
solutions satisfying the cluster property, we use eq.\,(\ref{vev}) and
require
\EQ
\frac{F_1^\varphi}{\lim_{\th\goto\infty}F_{11}^\varphi(\th)}=
\frac{F_2^\varphi}{\lim_{\th\goto\infty}F_{12}^\varphi(\th)}\,\,,
\lab{combined}
\EN
with $F_1^\varphi$ and $F_2^\varphi$ also determined in function of
$z_\varphi$ using eq.\,(\ref{pole}).

There exist only two values of the parameter $z_\varphi$ satisfying
the last equation. One of them exactly coincides with the value of
$z_\sigma$ which had been determined in ref.\,\cite{immf} 
without any reference to the cluster property
\EQ
z_\sigma=\frac{2m_1^2+m_3m_7}{2m_1^2}=4.869840..\,\,\,\,.
\lab{zsigma}
\EN
As we said above, once this initial condition has been fixed, all the matrix
elements of $\sigma(x)$ can in principle be determined using the residue
equations only, without further use of the cluster property. All the
one-particle and several two-particle FF of $\sigma(x)$ are given 
in tables 2 and 4,
respectively \footnote{Of course, these results coincide 
with those of ref.\,\cite{immf}.}. It must be 
stressed that all the two-particle FF of $\sigma(x)$ computed in this
way {\em automatically} satisfy the cluster property which then should
be regarded as characteristic of the whole solution selected by the
initial condition (\ref{zsigma}) \footnote{In other words,
the ratio in eq.\,(\ref{vev}) 
is the same for any $a$ and $b$. Moreover, once the standard CFT
normalisation of the operator is adopted, the value obtained for 
$\langle\sigma\rangle$ exactly coincides
with the result provided by the thermodynamic Bethe ansatz \cite{TBA}.}.
The same pattern is observed for the solution arising from the other
value of $z_\varphi$ satisfying eq.\,(\ref{combined}), which we
identify with $z_\varepsilon$ \footnote{The analytic expression for
$z_\var$ we dispose at the moment looks complicated and uninspiring
and we prefer to quote only its numerical value.}
\EQ
z_\var=1.255585..\,\,\,\,.
\lab{zvar}
\EN
The one and two-particle FF corresponding to this initial condition
are contained in tables 3 and 5; they were used in
ref.\,\cite{nonint} in order to compute the corrections 
the energy spectrum of the theory (\ref{action}) undergoes under a
small thermal perturbation induced by the energy operator $\var(x)$.
The remarkable agreement observed in ref.\,\cite{nonint} between the
theoretical predictions and the data coming from a numerical
diagonalisation of the Hamiltonian strongly supports the conclusion
that the cluster property correctly selects the matrix elements of
both relevant scaling operators in the IMMF.

The results contained in tables 2-5 can be used in the spectral
representation (\ref{spec}) in order to write down the large
distance expansion of the correlators $G_{\sigma\sigma}(x)$,
$G_{\sigma\var}(x)$ and $G_{\var\var}(x)$. The leading infrared
contributions are simply
\EQ
G_{\Phi_1\Phi_2}(x)=\langle\Phi_1\rangle\langle\Phi_2\rangle +
\frac{1}{\pi}\sum_{a=1}^3 F_a^{\Phi_1}F_a^{\Phi_2}K_0(m_a|x|) +
{\cal O}\left(e^{-2 m_1|x|}\right)
\label{threek}
\EN
where $K_0(x)$ is the modified Bessel function. In ref.\,\cite{immf},
the fast convergence of the FF series for $G_{\sigma\sigma}(x)$ was
tested against the Monte Carlo data available for that correlator; 
as far as we know, no similar data exist for
$G_{\sigma\var}(x)$ and $G_{\var\var}(x)$ but it seems natural to expect
a similar convergence pattern. An integral check is provided by the
following sum rule for the scaling dimension \cite{dsc}
\EQ
\Delta^\Phi=-\frac{1}{4\pi\langle\Phi\rangle}\int d^2x\langle\Theta(x)
\Phi(0)\rangle_c\,\,\,.
\lab{delta}
\EN
The results obtained for $\Delta^\sigma$ and $\Delta^\var$ using in
the last formula the spectral representation of the correlators
$G_{\sigma\sigma}(x)$ and $G_{\sigma\var}(x)$ are contained in tables
6 and 7, respectively ($\Delta^\Phi_{a_1\ldots a_n}$ denotes the
contribution coming from the intermediate state containing the
particles $A_{a_1}\ldots A_{a_n}$). We recall that the exact results
are $\Delta^\sigma=0.0625$ and $\Delta^\var=0.5$. The reason for the
slower convergence of the series for $\Delta^\var$ appears quite
clear. Indeed, according to (\ref{uv}), $G_{\sigma\var}(x)$ is much more
singular than $G_{\sigma\sigma}(x)$ as $x\goto 0$. As a consequence,
the integral in eq.\,(\ref{delta}) receives a larger contribution from
short distances for $\Phi=\var$ than for $\Phi=\sigma$. On the other
hand more and more terms in the spectral series are nedeed in order to
approximate precisely the correlators at small $x$.

In conclusion, it is clear that it would be higly desirable to reach a
satisfactory physical understanding of one of the basic ingredients we
used in this letter, namely the cluster property
(\ref{cluster}). It is very tempting to argue that this property (or
better, a suitable generalisation applying also to theories with
internal symmetries) characterises the matrix elements
of the scaling fields in two-dimensional quantum field theories. The
Smirnov's observation that, if the FF of an operator $\Phi(x)$
factorise asymptotically as in (\ref{cluster}), then it is
particularly simple to show that the correlator $G_{\Phi\Phi}(x)$
behaves as a power law at short distances \cite{SmirnovSG}, seems to 
go in this
direction. We hope that the results of this letter will stimulate
further investigations on this point.

\vspace{1cm}
{\em Acknowledgements.} We thank J.L. Cardy and G. Mussardo for
interesting discussions. G.D. was supported by the EPSRC grant
GR/J78044; P.S. was supported by a HEFCW grant.

\newpage
\hs
\vspace{25mm}

{\bf Table Caption}

\vspace{1cm}
\begin{description}
\item [Table 1]. Two-particle scattering amplitudes of the IMMF. 
Each factor $(\gamma)^\mu$ stays for $\left[\tanh\frac{1}{2}
\left(\th+i\pi \frac{\gamma}{30}\right)/\tanh\frac{1}{2}
\left(\th-i\pi \frac{\gamma}{30}\right)\right]^\mu$. 
The indeces $i$ placed above the functions $(\gamma)$ correspond to
the particles $A_i$ appearing as bound states in the $ab$ channel.
\item [Table 2]. One-particle Form Factors of the operator
$\sigma(x)$. The results are given in units of $m_1^{1/8}$ and refer
to the normalisation of the operator in which
$\langle\sigma\rangle=m_1^{1/8}$.
\item [Table 3]. One-particle Form Factors of the operator $\var(x)$. 
The results are given in units of $m_1$ and refer
to the normalisation of the operator in which
$\langle\var\rangle=m_1$.
\item [Table 4]. Coefficients of the polynomials $Q^\sigma_{ab}(\th)$.
The results are given in units of $m_1^{1/8}$ and refer
to the normalisation of the operator in which
$\langle\sigma\rangle=m_1^{1/8}$.
\item [Table 5]. Coefficients of the polynomials $Q^\var_{ab}(\th)$.
The results are given in units of $m_1$ and refer
to the normalisation of the operator in which
$\langle\var\rangle=m_1$.
\item [Table 6]. The first eight contributions to the sum rule for 
$\Delta^\sigma$.
\item [Table 7]. The first eight contributions to the sum rule for 
$\Delta^\var$.

\end{description}

%Beginning of tables

\newpage
\begin{center}

\vspace{3cm}
\begin{tabular}{|c|c|}\hline
$a$ \,\, $b$ &
$S_{ab}$ \\ \hline \hline
1 \,\, 1 &
$ \st{\bf 1}{(20)} \, \st{\bf 2}{(12)} \, \st{\bf 3}{(2)} $\\ \hline
1 \,\, 2 &
$ \st{\bf 1}{(24)} \, \st{\bf 2}{(18)} \, \st{\bf 3}{(14)} \, \st{\bf 4}{(8)}
$\\ \hline
1 \,\, 3 &
$ \st{\bf 1}{(29)} \, \st{\bf 2}{(21)} \, \st{\bf 4}{(13)} \,
\st{\bf 5}{(3)} \, (11)^2 $ \\ \hline
1 \,\, 4 &
$ \st{\bf 2}{(25)} \, \st{\bf 3}{(21)} \, \st{\bf 4}{(17)} \,
\st{\bf 5}{(11)} \, \st{\bf 6}{(7)} \, (15) $ \\ \hline
1 \,\, 5 &
$ \st{\bf 3}{(28)} \, \st{\bf 4}{(22)} \, \st{\bf 6}{(14)} \,
\st{\bf 7}{(4)} \, (10)^2 \, (12)^2 $ \\ \hline
1 \,\, 6 &
$ \st{\bf 4}{(25)} \, \st{\bf 5}{(19)} \, \st{\bf 7}{(9)} \,
(7)^2 \, (13)^2 \, (15) $ \\ \hline
1 \,\, 7 &
$ \st{\bf 5}{(27)} \, \st{\bf 6}{(23)} \, \st{\bf 8}{(5)} \,
(9)^2 \, (11)^2\, (13)^2 \, (15) $ \\ \hline
1 \,\, 8 &
$ \st{\bf 7}{(26)} \, \st{\bf 8}{(16)^3} \, (6)^2 \,
(8)^2 \, (10)^2 \, (12)^2 $ \\ \hline
2 \,\, 2 &
$ \st{\bf 1}{(24)} \, \st{\bf 2}{(20)} \, \st{\bf 4}{(14)} \,
\st{\bf 5}{(8)} \,\st{\bf 6}{(2)} \, (12)^2 $ \\ \hline
2 \,\, 3 &
$ \st{\bf 1}{(25)} \, \st{\bf 3}{(19)} \, \st{\bf 6}{(9)} \, (7)^2 \,
(13)^2 \, (15) $ \\ \hline
2 \,\, 4 &
$ \st{\bf 1}{(27)} \,\st{\bf 2}{(23)} \, \st{\bf 7}{(5)} \,
(9)^2 \, (11)^2 \, (13)^2\,(15)$ \\ \hline
2 \,\, 5 &
$ \st{\bf 2}{(26)} \,\st{\bf 6}{(16)^3} \, (6)^2 (8)^2 (10)^2 (12)^2 $
\\ \hline
2 \,\, 6 &
$ \st{\bf 2}{(29)} \, \st{\bf 3}{(25)} \, \st{\bf 5}{(19)^3} \,
\st{\bf 7}{(13)^3} \, \st{\bf 8}{(3)} \, (7)^2 (9)^2 (15) $ \\ \hline
2 \,\,7 &
$ \st{\bf 4}{(27)} \, \st{\bf 6}{(21)^3} \, \st{\bf 7}{(17)^3} \,
\st{\bf 8}{(11)^3} \, (5)^2 (7)^2 (15)^2 $ \\ \hline
2 \,\, 8 &
$ \st{\bf 6}{(28)} \, \st{\bf 7}{(22)^3} \, (4)^2 (6)^2 (10)^4 (12)^4 (16)^4 $
\\ \hline
3 \,\, 3 &
$ \st{\bf 2}{(22)} \, \st{\bf 3}{(20)^3} \, \st{\bf 5}{(14)} \,
\st{\bf 6}{(12)^3} \, \st{\bf 7}{(4)} \, (2)^2 $ \\ \hline
3 \,\, 4 &
$ \st{\bf 1}{(26)} \, \st{\bf 5}{(16)^3} \, (6)^2 (8)^2 (10)^2 (12)^2 $
\\ \hline
3 \,\, 5 &
$ \st{\bf 1}{(29)} \, \st{\bf 3}{(23)} \, \st{\bf 4}{(21)^3} \,
\st{\bf 7}{(13)^3} \, \st{\bf 8}{(5)} \, (3)^2 (11)^4 (15) $ \\ \hline
3 \,\, 6 &
$ \st{\bf 2}{(26)} \, \st{\bf 3}{(24)^3} \, \st{\bf 6}{(18)^3} \,
\st{\bf 8}{(8)^3} \, (10)^2 (16)^4 $ \\ \hline
3 \,\, 7 &
$ \st{\bf 3}{(28)} \, \st{\bf 5}{(22)^3} \, (4)^2 (6)^2 (10)^4 (12)^4 (16)^4
$ \\ \hline
3 \,\, 8 &
$ \st{\bf 5}{(27)} \, \st{\bf 6}{(25)^3} \, \st{\bf 8}{(17)^5} \,
(7)^4 (9)^4 (11)^2 (15)^3 $ \\ \hline
\end{tabular}
\end{center}

\begin{center}
Continued
\end{center}

\newpage
\begin{center}

\vspace{3cm}
\begin{tabular}{|c|c|}\hline
$a$ \,\, $b$ &
$S_{ab}$ \\ \hline \hline
4 \,\, 4 &
$ \st{\bf 1}{(26)} \, \st{\bf 4}{(20)^3} \, \st{\bf 6}{(16)^3} \,
\st{\bf 7}{(12)^3} \, \st{\bf 8}{(2)} \, (6)^2 (8)^2 $ \\ \hline
4 \,\, 5 &
$ \st{\bf 1}{(27)} \, \st{\bf 3}{(23)^3} \, \st{\bf 5}{(19)^3} \,
\st{\bf 8}{(9)^3} \, (5)^2 (13)^4 (15)^2 $ \\ \hline
4 \,\, 6 &
$ \st{\bf 1}{(28)} \, \st{\bf 4}{(22)^3} (4)^2 (6)^2 (10)^4
(12)^4 (16)^4 $ \\ \hline
4 \,\, 7 &
$ \st{\bf 2}{(28)} \, \st{\bf 4}{(24)^3} \, \st{\bf 7}{(18)^5} \,
\st{\bf 8}{(14)^5} \, (4)^2 (8)^4 (10)^4 $ \\ \hline
4 \,\, 8 &
$ \st{\bf 4}{(29)} \, \st{\bf 5}{(25)^3} \, \st{\bf 7}{(21)^5} \,
(3)^2 (7)^4 (11)^6 (13)^6 (15)^3 $ \\ \hline
5 \,\, 5 &
$ \st{\bf 4}{(22)^3} \, \st{\bf 5}{(20)^5} \, \st{\bf 8}{(12)^5} \,
(2)^2 (4)^2 (6)^2 (16)^4 $ \\ \hline
5 \,\, 6 &
$ \st{\bf 1}{(27)} \, \st{\bf 2}{(25)^3} \, \st{\bf 7}{(17)^5} \,
(7)^4 (9)^4 (11)^4 (15)^3 $ \\ \hline
5 \,\, 7 &
$ \st{\bf 1}{(29)} \, \st{\bf 3}{(25)^3} \, \st{\bf 6}{(21)^5} \,
(3)^2 (7)^4 (11)^6 (13)^6 (15)^3 $ \\ \hline
5 \,\, 8 &
$ \st{\bf 3}{(28)} \, \st{\bf 4}{(26)^3} \, \st{\bf 5}{(24)^5} \,
\st{\bf 8}{(18)^7} \, (8)^6 (10)^6 (16)^8 $ \\ \hline
6 \,\, 6 &
$ \st{\bf 3}{(24)^3} \, \st{\bf 6}{(20)^5} \, \st{\bf 8}{(14)^5} \,
(2)^2 (4)^2 (8)^4 (12)^6 $ \\ \hline
6 \,\, 7 &
$ \st{\bf 1}{(28)} \, \st{\bf 2}{(26)^3} \, \st{\bf 5}{(22)^5} \,
\st{\bf 8}{(16)^7} \, (6)^4 (10)^6 (12)^6 $ \\ \hline
6 \,\, 8 &
$ \st{\bf 2}{(29)} \, \st{\bf 3}{(27)^3} \, \st{\bf 6}{(23)^5} \,
\st{\bf 7}{(21)^7} \, (5)^4 (11)^8 (13)^8 (15)^4 $ \\ \hline
7 \,\, 7 &
$ \st{\bf 2}{(26)^3} \, \st{\bf 4}{(24)^5} \, \st{\bf 7}{(20)^7} \,
(2)^2 (8)^6 (12)^8 (16)^8 $ \\ \hline
7 \,\, 8 &
$ \st{\bf 1}{(29)} \, \st{\bf 2}{(27)^3} \, \st{\bf 4}{(25)^5} \,
\st{\bf 6}{(23)^7} \, \st{\bf 8}{(19)^9} \, (9)^8 (13)^{10} (15)^ 5 $
\\ \hline
8 \,\, 8 &
$ \st{\bf 1}{(28)^3} \, \st{\bf 3}{(26)^5} \, \st{\bf 5}{(24)^7} \,
\st{\bf 7}{(22)^9} \, \st{\bf 8}{(20)^{11}} \, (12)^{12} (16)^{12} $ \\
\hline
\end{tabular}
\end{center}

\begin{center}
{\bf Table 1}
\end{center}

\newpage

\begin{center}

\vspace{3cm}
\begin{tabular}{|c|}\hline
$ F^{\sigma}_1 =-0.64090211 $ \\
$ F^{\sigma}_2 =\,\,\,\, 0.33867436 $ \\
$ F^{\sigma}_3 =-0.18662854 $ \\
$ F^{\sigma}_4 =\,\,\,\, 0.14277176 $ \\
$ F^{\sigma}_5 =\,\,\,\, 0.06032607 $ \\
$ F^{\sigma}_6 =-0.04338937 $ \\
$ F^{\sigma}_7 =\,\,\,\, 0.01642569 $ \\
$ F^{\sigma}_8 =-0.00303607 $ \\ \hline
\end{tabular}
\end{center}

\begin{center}
{\bf Table 2}
\end{center}

\begin{center}

\vspace{3cm}
\begin{tabular}{|c|}\hline
$ F^{\var}_1 =-3.70658437 $ \\
$ F^{\var}_2 =\,\,\,\, 3.42228876 $ \\
$ F^{\var}_3 =-2.38433446 $ \\
$ F^{\var}_4 =\,\,\,\, 2.26840624 $ \\
$ F^{\var}_5 =\,\,\,\, 1.21338371 $ \\
$ F^{\var}_6 =-0.96176431 $ \\
$ F^{\var}_7 =\,\,\,\, 0.45230320 $ \\
$ F^{\var}_8 =-0.10584899 $ \\ \hline
\end{tabular}
\end{center}

\begin{center}
{\bf Table 3}
\end{center}

\newpage

\begin{center}

\vspace{3cm}
\begin{tabular}{|c|}\hline

$ c_{11}^1 = -2.093102832 $ \\
$ c_{11}^0 = -10.19307727 $ \\ \hline
$ c_{12}^2 = -7.979022182 $ \\
$ c_{12}^1 = -71.79206351 $ \\
$ c_{12}^0 = -70.29218939 $ \\ \hline
$ c_{13}^3 = -582.2557366 $ \\
$ c_{13}^2 = -6944.416956 $ \\
$ c_{13}^1 = -13406.48877 $ \\
$ c_{13}^0 = -7049.622303 $ \\ \hline
$ c_{22}^3 = -21.48559881 $ \\
$ c_{22}^2 = -333.8125724 $ \\
$ c_{22}^1 = -791.3745549 $ \\
$ c_{22}^0 = -500.2535896 $ \\ \hline
$ c_{14}^3 = 22.57778351 $ \\
$ c_{14}^2 = 318.7122159 $ \\
$ c_{14}^1 = 672.2210098 $ \\
$ c_{14}^0 = 377.4586311 $ \\ \hline
$ c_{15}^4 = -260.7643072 $ \\
$ c_{15}^3 = -4719.877128 $ \\
$ c_{15}^2 = -15172.07643 $ \\
$ c_{15}^1 = -17428.22924 $ \\
$ c_{15}^0 = -6716.787925 $ \\ \hline
$ c_{23}^4 = -92.73452350 $ \\
$ c_{23}^3 = -1846.579035 $ \\
$ c_{23}^2 = -6618.297073 $ \\
$ c_{23}^1 = -8436.850082 $ \\
$ c_{23}^0 = -3579.556465 $ \\ \hline
$ c_{33}^5 = -1197.056497 $ \\
$ c_{33}^4 = -30166.99117 $ \\
$ c_{33}^3 = -150512.4122 $ \\
$ c_{33}^2 = -301093.9432 $ \\
$ c_{33}^1 = -267341.1276 $ \\
$ c_{33}^0 = -87821.70785 $ \\ \hline
\end{tabular}
\end{center}

\begin{center}
Continued
\end{center}

\newpage
\begin{center}

\vspace{3cm}
\begin{tabular}{|c|}\hline
$ c_{25}^6 = 1425.995027 $ \\
$ c_{25}^5 = 44219.03877 $ \\
$ c_{25}^4 = 286184.1535 $ \\
$ c_{25}^3 = 788413.2178 $ \\
$ c_{25}^2 = 1078996.488 $ \\
$ c_{25}^1 = 725356.4417 $ \\
$ c_{25}^0 = 191383.5734 $ \\ \hline
$ c_{17}^5 = 190.8548023 $ \\
$ c_{17}^4 = 4633.706068 $ \\
$ c_{17}^3 = 21406.72691 $ \\
$ c_{17}^2 = 39514.82959 $ \\
$ c_{17}^1 = 32456.91939 $ \\
$ c_{17}^0 = 9906.265607 $ \\ \hline
$ c_{44}^7 = -7249.785565 $ \\
$ c_{44}^6 = -276406.7236 $ \\
$ c_{44}^5 = -2299573.212 $ \\
$ c_{44}^4 = -849276.3526 $ \\
$ c_{44}^3 = -16615618.39 $ \\
$ c_{44}^2 = -17950817.11 $ \\
$ c_{44}^1 = -10139089.36 $ \\
$ c_{44}^0 = -2341590.241 $ \\ \hline
\end{tabular}

\end{center}

\begin{center}
{\bf Table 4}
\end{center}

\newpage

\begin{center}

\vspace{3cm}
\begin{tabular}{|c|}\hline

$ c_{11}^1 = -70.00917205 $ \\
$ c_{11}^0 = -87.90247670 $ \\
\hline
$ c_{12}^2 = -466.3008246 $ \\
$ c_{12}^1 = -1307.331521 $ \\
$ c_{12}^0 = -853.2803886 $ \\
\hline
$ c_{13}^3 = -43021.45153 $ \\
$ c_{13}^2 = -182413.2733 $ \\
$ c_{13}^1 = -241929.7678 $ \\
$ c_{13}^0 = -102574.1349 $ \\
\hline
$ c_{22}^3 = -2193.896354 $ \\
$ c_{22}^2 = -10870.05277 $ \\
$ c_{22}^1 = -16161.44508 $ \\
$ c_{22}^0 = -7510.235388 $ \\
\hline
$ c_{14}^3 = 2074.636471 $ \\
$ c_{14}^2 = 9881.413381 $ \\
$ c_{14}^1 = 14357.04570 $ \\
$ c_{14}^0 = 6568.762583 $ \\
\hline
$ c_{15}^4 = -30333.56619 $ \\
$ c_{15}^3 = -198757.2340 $ \\
$ c_{15}^2 = -447504.5720 $ \\
$ c_{15}^1 = -422808.9295 $ \\
$ c_{15}^0 = -143743.2050 $ \\
\hline
$ c_{23}^4 = -11971.94909 $ \\
$ c_{23}^3 = -81253.72269 $ \\
$ c_{23}^2 = -186593.8661 $ \\
$ c_{23}^1 = -178494.3378 $ \\
$ c_{23}^0 = -61194.62416 $ \\
\hline
$ c_{33}^5 = -195385.7662 $ \\
$ c_{33}^4 = -1743171.802 $ \\
$ c_{33}^3 = -5603957.324 $ \\
$ c_{33}^2 = -8422606.859 $ \\
$ c_{33}^1 = -6035102.896 $ \\
$ c_{33}^0 = -1668721.004 $ \\
\hline
\end{tabular}
\end{center}

\begin{center}
Continued
\end{center}

\newpage
\begin{center}

\vspace{3cm}
\begin{tabular}{|c|}\hline

$ c_{25}^6 = 289831.4882 $ \\
$ c_{25}^5 = 3275586.983 $ \\
$ c_{25}^4 = 13872077.63 $ \\
$ c_{25}^3 = 29236961.96 $ \\
$ c_{25}^2 = 32979257.31 $ \\
$ c_{25}^1 = 19100224.04 $ \\
$ c_{25}^0 = 4471623.121 $ \\
\hline
$ c_{17}^5 = 30394.23374 $ \\
$ c_{17}^4 = 274294.8033 $ \\
$ c_{17}^3 = 897781.3229 $ \\
$ c_{17}^2 = 1.375919456 $ \\
$ c_{17}^1 = 1.004969466 $ \\
$ c_{17}^0 = 282938.1974 $ \\
\hline
$ c_{44}^7 = -1830120.693 $ \\
$ c_{44}^6 = -25699492.93 $ \\
$ c_{44}^5 = -138411873.8 $ \\
$ c_{44}^4 = -384776478.8 $ \\
$ c_{44}^3 = -608371427.1 $ \\
$ c_{44}^2 = -553818699.0 $ \\
$ c_{44}^1 = -270964337.7 $ \\
$ c_{44}^0 = -55283137.91 $ \\
\hline
\end{tabular}

\end{center}

\begin{center}
{\bf Table 5}
\end{center}

\newpage

\, \, \,
\vspace{3cm}

\begin{center}

\vspace{3cm}
\begin{tabular}{|l|c|}\hline

$\Delta^\sigma_1 $ &   0.0507107  \\
$\Delta^\sigma_2 $ &   0.0054088 \\
$\Delta^\sigma_3 $ &   0.0010868 \\
$\Delta^\sigma_{11}$ & 0.0025274 \\
$\Delta^\sigma_4 $ &   0.0004351 \\
$\Delta^\sigma_{12}$ & 0.0010446 \\
$\Delta^\sigma_5 $ &   0.0000514 \\
$\Delta^\sigma_{13}$ & 0.0002283 \\ \hline
$\Delta^\sigma_{\rm partial}$ & 0.0614934 \\ \hline
\end{tabular}
\end{center}
\vspace{1cm}

\begin{center}
{\bf Table 6}
\end{center}

\begin{center}

\vspace{3cm}
\begin{tabular}{|l|c|}\hline

$\Delta^\var_1 $ &   0.2932796 \\
$\Delta^\var_2 $ &   0.0546562 \\
$\Delta^\var_3 $ &   0.0138858 \\
$\Delta^\var_{11}$ & 0.0425125 \\
$\Delta^\var_4 $ &   0.0069134 \\
$\Delta^\var_{12}$ & 0.0245129 \\
$\Delta^\var_5 $ &   0.0010340 \\
$\Delta^\var_{13}$ & 0.0065067 \\ \hline
$\Delta^\var_{\rm partial}$ & 0.4433015 \\ \hline
\end{tabular}
\end{center}
\vspace{1cm}

\begin{center}
{\bf Table 7}
\end{center}

\end{document}